\documentclass[pra,twocolumn,showpacs,superscriptaddress,floatfix]{revtex4}
\usepackage[dvips]{graphicx}
\newcommand{\Cdot}{{\,{\cdot}\,}}
\newcommand{\IXY}{I(X\,{:}\,Y)}
\newcommand{\keto}{\left|0\right\rangle}
\newcommand{\ketpsiio}{\left|{\psi_{i_1}}\right\rangle}
\newcommand{\ketpsiin}{\left|{\psi_{i_n}}\right\rangle}
\newcommand{\ketpsii}{\left|{\psi_i}\right\rangle}
\newcommand{\ketpsij}{\left|{\psi_j}\right\rangle}
\newcommand{\brapsii}{\left\langle{\psi_i}\right|}
\newcommand{\brapsij}{\left\langle{\psi_j}\right|}
\newcommand{\ketpsikpm}{|{\psi_{k_{\pm}}}\rangle}
\newcommand{\ketpsjithetao}{\left|\psi'_i(\theta_0)\right\rangle}
\newcommand{\ketaj}{\left|{a_j}\right\rangle}
\newcommand{\braaj}{\left\langle{a_j}\right|}
\newcommand{\braomegaj}{\left\langle{\omega_j}\right|}
\newcommand{\ketomegaj}{\left|{\omega_j}\right\rangle}
\newcommand{\ketomegaz}{\left|{\omega_0}\right\rangle}
\newcommand{\ketomegao}{\left|{\omega_1}\right\rangle}
\newcommand{\ketomegat}{\left|{\omega_2}\right\rangle}
\newcommand{\ketpsiz}{\left|{\psi_0}\right\rangle}
\newcommand{\ketpsio}{\left|{\psi_1}\right\rangle}
\newcommand{\ketpsit}{\left|{\psi_2}\right\rangle}
\newcommand{\ketpsih}{\left|{\psi_3}\right\rangle}
\newcommand{\ketpsif}{\left|{\psi_4}\right\rangle}
\newcommand{\ketpsiv}{\left|{\psi_5}\right\rangle}
\newcommand{\ketpsis}{\left|{\psi_6}\right\rangle}
\newcommand{\ketH}{{\left|\leftrightarrow\right\rangle}}
\newcommand{\ketV}{{\left|\,\updownarrow\,\right\rangle}}
\newcommand{\Hsp}{\mathcal{H}_s}
\newcommand{\aiH}{A_\mathrm{H}}
\newcommand{\aiV}{A_\mathrm{V}}
\newcommand{\biH}{B_\mathrm{H}}
\newcommand{\biV}{B_\mathrm{V}}
\newcommand{\aoH}{A'_\mathrm{H}}
\newcommand{\aoV}{A'_\mathrm{V}}
\newcommand{\boH}{B'_\mathrm{H}}
\newcommand{\boV}{B'_\mathrm{V}}
\newcommand{\affA}{%
    Communications Research Laboratory,
    Koganei, Tokyo 184-8795, Japan}
\newcommand{\affB}{%
    CREST, Japan Science and Technology Corporation}
\newcommand{\affC}{%
    Department of Physics and Applied Physics,
    University of Strathclyde, Glasgow G4 0NG, Scotland}
\begin{document}
\title{%
    Optimum detection for extracting maximum information
    from symmetric qubit sets}
\author{Jun Mizuno}
\affiliation{\affA}
\affiliation{\affB}
\author{Mikio Fujiwara}
\affiliation{\affA}
\affiliation{\affB}
\author{Makoto Akiba}
\affiliation{\affA}
\author{Tetsuya Kawanishi}
\affiliation{\affA}
\author{Stephen M. Barnett}
\affiliation{\affC}
\author{Masahide Sasaki}
\affiliation{\affA}
\affiliation{\affB}
\email{e-mail:psasaki@crl.go.jp}
\begin{abstract}
We demonstrate a class of optimum detection strategies for
extracting the maximum information from sets of equiprobable
real symmetric qubit states of a single photon.
These optimum strategies have been predicted by
Sasaki \textit{et al.}~\cite{SasakiBarnettJozsa99}.
The peculiar aspect is that the detections with
at least three outputs suffice for optimum extraction
of information regardless of the number of signal elements.
The cases of ternary (or trine), quinary, and septenary
polarization signals are studied where a standard von Neumann
detection (a projection onto a binary orthogonal basis)
fails to access the maximum information.
Our experiments demonstrate
that it is possible with present technologies
to attain about 96\,\% of the theoretical limit.
\end{abstract}
\pacs{03.67.Hk, 03.65.Ta, 42.50.--p}
%
\date{September 5, 2001}
%
\maketitle
%
%
\section{Introduction}
\label{sc:intro}
In communications systems a sender, Alice, represents \textit{messages},
for example the alphabet $\{a, b, \ldots, z\}$,
by a given set of \textit{letters} $\{x_i\}$ such as $\{0,1\}$.
She transmits sequences of the letters, in the form of
\textit{codewords}, through a communication channel.
A receiver, Bob, detects codewords and thereby retrieves the message.
To design an optimum communication system, one should first
know basic properties of distinguishing the letter set $\{x_i\}$
over a channel.
These are specified by conditional probabilities $P(y_j\vert x_i)$
that Bob finds the outcome $y_j$ when Alice selected the letter~$x_i$.
The matrix of these conditional probabilities, $[P(y_j\vert x_i)]$,
is called the channel matrix.
All the physical properties of the channel and of the detector
are modeled through this channel matrix.

When a communication system operates in quanta,
the channel matrix will be determined by the rules of quantum mechanics.
The physical carrier conveying a letter $x_i$ should explicitly be
described by a quantum state $\ketpsii$, which we refer to
as the \textit{letter state}.
For example, if the letters $\{0, 1\}$ are conveyed by weak pulses
of laser light, the corresponding quantum states $\{\,\ketpsii\}$
are usually nonorthogonal coherent states.
Such nonorthogonal states can never be distinguished perfectly,
even in principle.
Therefore even if a channel and a detector are completely noiseless,
quantum mechanics imposes an inevitable source of error or ambiguity
in signal detection.

A detection process is represented mathematically
by the probability operator measure (POM),
which consists of nonnegative (generally not normalized)
Hermitian operators satisfying the resolution of the identity
\cite{Helstrom_QDET,Holevo_book,Peres_book}:
\begin{equation}
\widehat\Pi_j^\dagger=\widehat\Pi_j,\quad
\widehat\Pi_j\geq 0\quad{}^{\forall} j, \quad
\sum_j\widehat\Pi_j=\hat I.
\label{def_POM}
\end{equation}
Each element $\widehat\Pi_j$ is associated with the measurement
outcome $j$ and hence implies the output letter $y_j$.
If a channel is noiseless and hence quantum limited,
then the channel matrix is given by
\begin{equation}
P(y_j\vert x_i) =
  \brapsii\widehat\Pi_j\ketpsii
.
\label{P(y|x)}
\end{equation}

The primary concern in quantum communication is to determine
the optimum detection strategy $\{\widehat\Pi_j\}$
to distinguish among the letter states $\{\,\ketpsii\}$.
Each state $\ketpsii$ encodes the classical information
embodied in the classical letter $x_i$,
which is selected with known prior probability~$\{P(x_i)\}$.

The meaning of `optimum' depends on a task that we are going to do.
The simplest requirement is that Bob wants to decide which
letter state he has received among the set $\{\,\ketpsii\}$
with the smallest error.
This usually means minimizing the average error probability,
or bit error rate $P_{\mathrm{e}}$
\cite{Holevo73_condition,YuenKennedyLax75}.
A second possibility is for Bob to eliminate all errors
by allowing the possibility of inconclusive results
by means of unambiguous state discrimination
\cite{Ivanovic87,Dieks88,Peres88,Jaeger95,Barnett97RSL,%
      CheflesBarnett98,Chefles98_USD,CheflesContemp}.
The optimum strategy in this case will be the one that minimizes
the probability~$P_{\mathrm{i}}$ of inconclusive outcomes.
This type of detection has been proposed for
quantum key distribution~\cite{HuttnerPeres}.

For the communication of messages, however, Bob does best by devising
a detection strategy so as to retrieve Alice's message with the
greatest probability.
This does not necessarily mean minimizing
either $P_{\mathrm{e}}$ or $P_{\mathrm{i}}$, but instead
means reducing the uncertainty in some
\textit{random variable} $X=\{ x_i, P(x_i) \}$.
Such a detection strategy is directly related to reliable
communication by coding technique and
is actually used as a basic building block for
effective decoding procedures of codeword states formed from the
letter states $\{\,\ketpsii\}$.
(A more detailed explanation
of this point is given in Appendix.)  

The reduction of the uncertainty caused by a detection
is quantified by the Shannon mutual information $\IXY$
between the input (Alice's) and output (Bob's) random variables,
$X=\{ x_i, P(x_i) \}$ and $Y=\{ y_j, P(y_j) \}$.
This mutual information $\IXY$ can be regarded as the amount
of information extracted from $X$.
Bob's optimum strategy
will be the one that maximizes $\IXY$.
Other figures of merit
have also been considered and these include the fidelity
\cite{Schumacher95,Claire01}.

The optimum conditions are already known for minimizing
the error probability
\cite{Holevo73_condition,YuenKennedyLax75}.
It is not an easy task, however, to find the optimum detection
strategies from these conditions.
In fact, optimum strategies are only known for some special cases
such as the set of binary states, sets of symmetric states
\cite{Holevo73_condition,YuenKennedyLax75,Osaki96_OptPOM,%
      BanKurokawa97_SqRt,Sasaki98a}
and multiply symmetric states \cite{Barnett01}.
Unambiguous state discrimination is possible if and only if
the letter states are linearly independent and an explicit
method for constructing the optimum strategy has been given in
this case \cite{Chefles98_USD,CheflesContemp}.
Finding optimum solutions for $\IXY$ is much more difficult
than those for $P_{\mathrm{e}}$ and $P_{\mathrm{i}}$
due to the nonlinearity of logarithmic function of $\IXY$
with respect to a POM\@.
Optimum solutions are known only for
the set of binary pure states
\cite{Levitin95_QCM94,Osaki2000_QCM98}
and for sets of real symmetric qubit states with equal
prior probabilities~\cite{Davies78,SasakiBarnettJozsa99}.

It seems intuitively reasonable that we might obtain
most information by minimizing either
the average error probability $P_{\mathrm{e}}$ or
the probability of inconclusive outcomes $P_{\mathrm{i}}$.
In fact, the maximum mutual information for binary states is
attained by the same strategy that realizes the minimum
average error probability.
There are, however, cases where the maximum information must
be obtained neither by minimizing $P_{\mathrm{e}}$ nor $P_{\mathrm{i}}$
\cite{Davies78,SasakiBarnettJozsa99,Shor2001}.

Devices capable of demonstrating near optimum detection at
the single photon level have been demonstrated in the laboratory.
The simplest of these is discrimination between the set of
binary photon polarization states with the minimum allowed
average error probability
\cite{Barnett97_exp}.
Unambiguous discrimination between two non-orthogonal polarization
states has also been demonstrated \cite{Huttner96a,Clarke00a}.
A set of more than three polarization states is linearly dependent
and hence it is not possible to carry out unambiguous
state discrimination.
Clarke \textit{et al}.\ have demonstrated state discrimination
with near minimum error probability for both the trine and
tetrad polarization states
\cite{Clarke01b}.
They have also demonstrated the ability to extract more
information than is possible by the best, standard von Neumann
measurement
(a projection onto binary orthogonal polarization states).

In this paper we describe our experimental implementation of
a class of optimum strategies for maximizing the mutual information,
as predicted by
Ref.\,\cite{SasakiBarnettJozsa99}.
One of these is the ternary or trine set of states discussed
by Clarke \textit{et al}.~\cite{Clarke01b}.
We have improved upon the information yield obtained by
these authors and have also measured the information obtained
from signals formed from five and seven possible polarizations.
Our letter states are implemented physically as single photon
polarizations.
The required equiprobable real symmetric qubit states
are then states of linear polarization.
Such sets of states have previously found application in
quantum key distribution
\cite{PhoenixTownsend95,PhoenixBarnettChefles00}.
From the view point of fundamental interests,
they might be the simplest system with which to test
the peculiar effect predicted by Davies' theorem.
According to the theorem, there must exist at least
one solution, that maximizes the mutual information,
which has $N$ possible outputs, where $N$ is bounded by
$d \le N \le d^2$
with $d$ being the dimension of the Hilbert space $\Hsp$
supported by Alice's set \cite{Davies78}.
For real state sets, this bounding inequality becomes
$d \le N \le d(d+1)/2$
\cite{SasakiBarnettJozsa99}.
Thus for a single photon polarization system,
one can always optimize the mutual information by constructing
a device with just three possible outputs.
This is true regardless of the number of letter states.
In the case of ternary or trine signals, the optimum measurement
consists of three symmetric state vectors with the length less
than the unity, and has been demonstrated experimentally in
Ref.\,\cite{Clarke01b}.
In the cases of quinary and septenary signals, the optimum strategies
consist of three nonorthogonal state vectors with different lengths.
In the septenary case, there are two different configurations
of measurement state vectors.
We study how each of these strategies work and the extent to which
they allow us to access the theoretical maximum amount of
mutual information.

%
%
\section{Real symmetric qubit sets and optimum detection}
\label{sc:sig$meas}
Let $\{\,\ketH, \ketV\}$ be the orthogonal basis
of linear polarization states of a single photon.
Then the real symmetric qubit states are defined as
\begin{eqnarray}
\ketpsii &=&
   \cos\frac{i\pi}{M} \ketH + \sin\frac{i\pi}{M} \ketV
\label{psi_i}\\
&& (i=0, \ldots , M{-}1).
\nonumber
\end{eqnarray}
We assume that each state is selected
with equal prior probability $1/M$.
This set is
one of the few quantum state sets
for which optimum strategies for the accessible information
are explicitly known
\cite{Levitin95_QCM94,Osaki2000_QCM98,Davies78,SasakiBarnettJozsa99}.

\begin{figure}
\begin{center}
\includegraphics[width=0.41\textwidth]{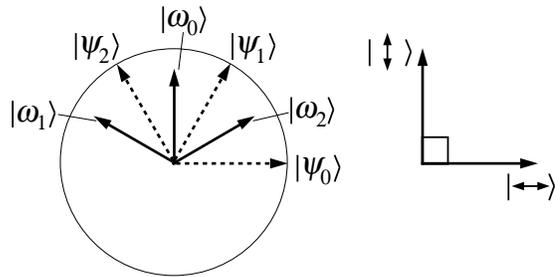}
\end{center}
\caption{\label{fg:3sig&w}
The measurement state vectors for the optimum strategy
(solid line) and the signal state vectors in the case of
the ternary (trine) signals.
}
\end{figure}
For $M>2$, the signal states cannot be distinguished perfectly,
thus $P_{\mathrm{i}} = 1$.
The minimum average error probability is
\begin{equation}
P_{\mathrm{e}}=1-{2\over M},
\end{equation}
which is attained by the POM $\{ \widehat{\Pi}_j \}$
\cite{Helstrom_QDET,BanKurokawa97_SqRt}
\begin{eqnarray}
\widehat{\Pi}_j &=& \ketaj \! \braaj
\qquad\text{with}
\\
\ketaj &=& \sqrt{2\over M}
\left(
   \cos\frac{j\pi}{M} \ketH + \sin\frac{j\pi}{M} \ketV
\right)
\label{a_j}\\
&&
(j=0, \ldots , M{-}1)
\,.
\nonumber
\end{eqnarray}
This POM is unique in leading to the minimum error probability
and has the same number of POM elements,
corresponding to the measurement outcomes, as the letter states.

In contrast, maximizing the mutual information requires
a POM with three rank-one elements at most,
corresponding to just three measurement outcomes~%
\cite{SasakiBarnettJozsa99}.
Although
it is also possible to construct optimum POMs with
elements more than three,
a strategy with minimum outputs
is often the one desired in practice.

If $M$ is even,
a von Neumann measurement, i.e.\ a pair of orthogonal projectors,
can be the optimum strategy with minimum outputs.
If $M$ is odd, then at least three outputs are required and
a standard von Neumann measurement fails
in maximizing the mutual information.
The three rank-one elements required for the optimum POM
$\{ \widehat{\Pi}_j \}$
are specified as follows:
\begin{eqnarray}
\widehat{\Pi}_j &=& \ketomegaj\!\braomegaj
\\
\text{with} &&
\left\{
\begin{array}{lcl}
\ketomegaz &=&
    - \sin{\gamma\over2}\,\ketV  \\[1\jot]
\ketomegao &=& {1\over{\sqrt2}} \left(
    - \ketH + \cos{\gamma\over2}\,\ketV   \right) \\[1\jot]
\ketomegat &=& {1\over{\sqrt2}} \left(
      \ketH + \cos{\gamma\over2}\,\ketV   \right)
\end{array}
\right.  
\label{vec_omega_m_n}
\end{eqnarray}
where $\gamma$ is determined from
\begin{equation}
\cos{\gamma\over2} \equiv \cot{{m\pi}\over M},
\quad
\sin{\gamma\over2} \equiv
-\sqrt{1-\cot^2{{m\pi}\over M}}
\label{angle_gamma}
\end{equation}
for an integer parameter $m$ within the range
${M\over4}<m<{M\over2}$.
We will refer to the unnormalized vectors given in
Eq.\,(\ref{vec_omega_m_n}) as measurement state-vectors.
\begin{table}
\begin{center}
\begin{tabular}{|l|l|l|l|}
\hline
 & $\ketpsiz$ & $\ketpsio$ & $\ketpsit$ \\
\hline
$\ketomegaz$ & 0   & 0.5 & 0.5 \\
$\ketomegao$ & 0.5 & 0   & 0.5 \\
$\ketomegat$ & 0.5 & 0.5 & 0   \\
\hline
\end{tabular}
\end{center}
\vspace{-1em}
\caption{\label{tb:TrineRatio}
The channel matrix of the optimum POM for the ternary signals.
}
\end{table}

In the case of $M=3$ (ternary or trine),
the optimum POM is given by $m=1$ which results in
the set of three measurement state-vectors with equal norms.
The signal and measurement state-vectors are
schematically shown in Fig.\,\ref{fg:3sig&w}.
In this figure,
each arrow represents the polarization direction
where the horizontal and the vertical directions
correspond to the two unit bases $\ketH$ and $\ketV$,
respectively.
The length of each arrow represents the norm of the associated
state vector, e.g.\ $\ketpsii$ or $\ketomegaj$.

The optimum measurement in this case means that
the state vectors $\ketpsij$ and $\ketomegaj$
are orthogonal, and thus
\begin{eqnarray}
P(y_j|x_j) &=& \brapsij \widehat\Pi_j\ketpsij = 0
\,.
\end{eqnarray}
The other two possible measurement outcomes occur
with equal probabilities.
This situation is summarized in Table~\ref{tb:TrineRatio}.

\begin{figure}
\begin{center}
\includegraphics[width=0.23\textwidth]{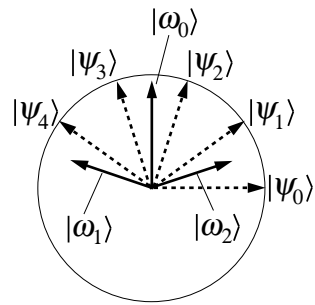}
\end{center}
\caption{\label{fg:5sig&w}
The measurement state vectors for the optimum strategy
(solid line) and the signal state vectors in the case
of the quinary signals ($M=5$).
}
\end{figure}

\begin{figure}[t]
\begin{center}
\includegraphics[width=0.48\textwidth]{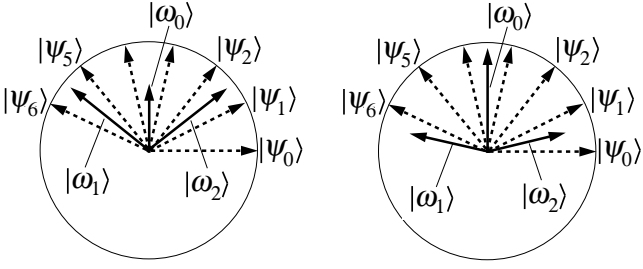}
\end{center}
\caption{\label{fg:7sig&w_m=2&3}
The two optimum strategies in the case of the septenary
signals ($M=7$).
The left figure corresponds to the choice $m=2$, while
the right one corresponds to the other choice $m=3$.
}
\end{figure}

In the cases of $M=5$ (quinary) and $M=7$ (septenary),
Eq.\,(\ref{vec_omega_m_n}) results in
the three measurement state-vector with two distinct norms.
The relationship between the quinary letter states and
the three measurement state-vectors (with $m=2$) is depicted
in Fig.\,\ref{fg:5sig&w}.
The channel matrix in this case is
summarized in Table~\ref{tb:QuinaryRatio}.
\begin{table}
\begin{center}
\begin{tabular}{|l|@{~}l|@{~}l|@{~}l|@{~}l|@{~}l|}
\hline
 & $\ketpsiz$ & $\ketpsio$ & $\ketpsit$ &
   $\ketpsih$ & $\ketpsif$ \\
\hline
$\ketomegaz$ & 0   & 0.309 & 0.809 & 0.809 & 0.309 \\
$\ketomegao$ & 0.5 & 0.191 & 0     & 0.191 & 0.5   \\
$\ketomegat$ & 0.5 & 0.5   & 0.191 & 0     & 0.191 \\
\hline
\end{tabular}
\end{center}
\vspace{-1em}
\caption{\label{tb:QuinaryRatio}
The channel matrix of the optimum POM for the quinary signals.
}
\end{table}
In the septenary case, there are two different POMs
with three elements given by Eq.\,(\ref{vec_omega_m_n}),
with $m=2$ and $m=3$ in Eq.\,(\ref{angle_gamma}) respectively.
They are depicted in Fig.\,\ref{fg:7sig&w_m=2&3}
and summarized in
Tables \ref{tb:SeptenaryRatio2} and~\ref{tb:SeptenaryRatio3}.
In either case, there are combinations of $(i,j)$ that
give $P(y_j|x_i)=0$, although $j$ is not necessarily
equal to $i$ (a difference from the ternary case).
\begin{table}
\begin{center}
\begin{tabular}{|l|@{~}l|@{~}l|@{~}l|@{~}l|@{~}l|@{~}l|@{~}l|}
\hline
 & $\ketpsiz$ & $\ketpsio$ & $\ketpsit$ & $\ketpsih$
 & $\ketpsif$ & $\ketpsiv$ & $\ketpsis$ \\
\hline
$\ketomegaz$ & 0   & 0.069 & 0.223 & 0.346 & 0.346 & 0.223 & 0.069 \\
$\ketomegao$ & 0.5 & 0.154 & 0     & 0.154 & 0.5   & 0.777 & 0.777 \\
$\ketomegat$ & 0.5 & 0.777 & 0.777 & 0.5   & 0.154 & 0     & 0.154 \\
\hline
\end{tabular}
\end{center}
\vspace{-1em}
\caption{\label{tb:SeptenaryRatio2}
The channel matrix of the optimum POM with $m{=}2$
(see Eq.\,(\ref{angle_gamma})\,) for the septenary signals.
}
\begin{center}
\begin{tabular}{|l|@{~}l|@{~}l|@{~}l|@{~}l|@{~}l|@{~}l|@{~}l|}
\hline
 & $\ketpsiz$ & $\ketpsio$ & $\ketpsit$ & $\ketpsih$
 & $\ketpsif$ & $\ketpsiv$ & $\ketpsis$ \\
\hline
$\ketomegaz$ & 0   & 0.178 & 0.579 & 0.901 & 0.901 & 0.579 & 0.178 \\
$\ketomegao$ & 0.5 & 0.322 & 0.099 & 0     & 0.099 & 0.322 & 0.5   \\
$\ketomegat$ & 0.5 & 0.5   & 0.322 & 0.099 & 0     & 0.099 & 0.322 \\
\hline
\end{tabular}
\end{center}
\vspace{-1em}
\caption{\label{tb:SeptenaryRatio3}
The channel matrix of the optimum POM with $m{=}3$
(see Eq.\,(\ref{angle_gamma})\,) for the septenary signals.
}
\end{table}

The method to implement
the optimum POM with minimum outputs,
as given in Eq.\,(\ref{vec_omega_m_n}),
is prescribed in detail in Ref.\,\cite{SasakiBarnettJozsa99}.
In short, the nonorthogonal measurement basis
$\{\,\ketomegaj\}$
is considered as the projection
of a three-dimensional orthonormal basis in an enlarged space.
Such an enlarged space is achieved by introducing
another independent binary basis.

\begin{figure}[t]
\begin{center}
\includegraphics[width=0.4\textwidth]{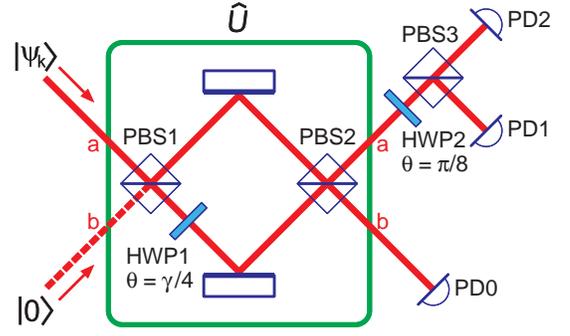}
\end{center}
\caption{\label{fg:pmzprin}
Principle of the detector that realizes
the optimum POM\@.
Here PBS stands for a polarizing beam splitter,
HWP for a half waveplate whose axis is rotated by
$\theta$, and PD for a photodetector.
}
\end{figure}

In practice, the concept described above is realized
as the polarization Mach--Zehnder interferometer
shown in Fig.\,\ref{fg:pmzprin}.
The four-dimensional space is composed of
$\{\,\ketH_a, \ketV_a, \ketH_b, \ketV_b \}$,
where subscripts represent the optical paths ($a,b$)
indicated in Fig.\,\ref{fg:pmzprin}.
Our letter states present in the subspace spanned
by the first two of these vectors.
The additional port (at $b$ in Fig.\,\ref{fg:pmzprin})
with an input of vacuum state $\keto$
enlarges the space.

The unitary operation of the Mach--Zehnder part
(indicated as $\widehat{U}$ in Fig.\,\ref{fg:pmzprin})
can be written as
\begin{eqnarray}
\lefteqn{
\aoH \ketH_{a} \!\!+ \aoV \ketV_{a} \!\!+
\boH \ketH_{b} \!\!+ \boV \ketV_{b}
}\nonumber\\
&=& \widehat{U} \bigl(
\aiH \ketH_{a} + \aiV \ketV_{a} + \biH \ketH_{b} + \biV \ketV_{b}
\bigr)
\\
\lefteqn{
\text{with }
\left[
\begin{array}{c}
\aoH \\ \aoV \\ \boH \\ \boV
\end{array}
\right]
=
\left[
\begin{array}{cccc}
1 & 0 & 0 & 0 \\
0 & \cos\gamma/2 & \sin\gamma/2 & 0 \\
0 &-\sin\gamma/2 & \cos\gamma/2 & 0 \\
0 & 0 & 0 & 1
\end{array}
\right]
\left[
\begin{array}{c}
\aiH \\ \aiV \\ \biH \\ \biV
\end{array}
\right]
}
\nonumber
\end{eqnarray}
where $\gamma/2$ is twice the angle of HWP1.
(This $\gamma/2$ represents the angle of one of the unit basis
in the enlarged space relative to the signal plane.)
In our setup, the inputs are $\biH=\biV=0$
and hence $\boV=0$.
Thus the apparatus of Fig.\,\ref{fg:pmzprin} actually
couples a three-dimensional state space.

PD0 detects $\ketH_{b}$ components whose amplitude
is given by
\begin{eqnarray}
\boH = -\sin(\gamma/2)\aiV
\,.
\label{eq:PD0}
\end{eqnarray}
Its null result guarantees that
the signal was not $\ketpsiz$.
On the other hand,
$\ketH_{a}$ and $\ketV_{a}$ components
are further mixed at HWP2 and PBS3, 
resulting in amplitudes of
\begin{eqnarray}
{1\over\sqrt{2}} \bigl( \aoH \pm \aoV \bigr)
&=& {1\over\sqrt{2}}
\left[ \aiH \pm \cos{\gamma\over2}\,\aiV \right]
\label{eq:PD12}
\end{eqnarray}
which are then detected at PD1 and PD2.
By inspecting Eqs.\,(\ref{eq:PD0}) and~(\ref{eq:PD12}),
it can be seen that
$\ketomegaj$ given in Eq.\,(\ref{vec_omega_m_n})
were reproduced.
When the condition Eq.\,(\ref{angle_gamma}) is satisfied,
the null result at PD1 or PD2 excludes
one of the possible signals
(\,$\ketpsikpm$ with $k_{+}=M{-}m$ and $k_{-}=m$).
%
%
\section{Experiment}\label{sc:exp}
\begin{figure}
\includegraphics[width=0.48\textwidth]{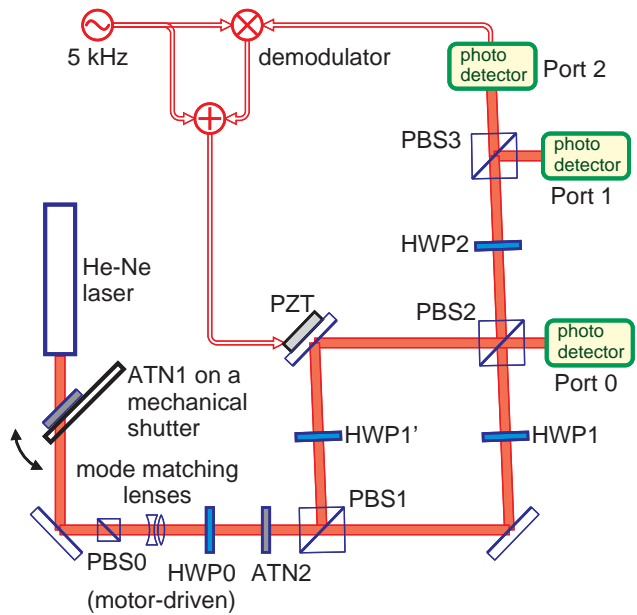}
\caption{\label{fg:layout}
Experimental configuration.
The same symbols as in Fig.\,\ref{fg:pmzprin}
and ATN for an attenuator are used.
Each of Ports 0, 1, and 2 contains an APD and
a silicon photodiode with a mechanical shutter to
switch the beam between them.
All PBS are adjusted for the maximum separation
of two polarization, resulting in a slightly
($\approx0.02$\,rad) slanted parallelogram arrangement
for the Mach--Zehnder.
}
\end{figure}

The principle described in the previous section is
realized in an actual setup to confirm
the theoretical results.
In the experiment, the polarization basis
$\{\,\ketH, \ketV \}$
correspond to P- (within the paper plane in Fig.\,\ref{fg:layout})
and S- (perpendicular to the paper plane) polarizations,
respectively.

The light source is a He--Ne laser (Spectra--Physics, model 117A)
operating at the wavelength of 632.8\,nm.
The laser light of 1\,mW is first attenuated by the attenuator
ATN1 by a factor of $10^{-6}$,
purified to the horizontally polarized state by
the polarizing beam splitter PBS0.
The half waveplate HWP0, driven by a stepping motor,
works as a modulator to produce the set $\{\,\ketpsii\}$.
Then the beam is further attenuated by ATN2
by a factor of $10^{-4}$.
At the input of the Mach--Zehnder interferometer,
the light power is of order $10^{-4}$\,fW
($\approx 3\Cdot 10^5$\,photon/sec).
In other words, the beam contains about $10^{-3}$ photons in
one meter,
whereas our detecting circuit is shorter than that.

The polarization Mach--Zehnder interferometer is composed
of two PBSs, PBS1 and PBS2.
Each PBS is carefully mounted so as to operate with
an extinction ratio of $1\,{:}\,1000$
(see below and Ref.\,\cite{pmzexp}).
Each path of the Mach--Zehnder contains one half waveplate,
HWP1 and HWP$1'$.
The angle of HWP1 is adjusted to a quarter of
$\gamma$ in Eq.\,(\ref{angle_gamma})
so that the polarization of the light is rotated by $\gamma/2$,
whereas HWP$1'$ is inserted for symmetry and thus adjusted
not to affect the polarization state.

The beams from the two paths are superimposed at PBS2,
resulting in two output beams from the Mach--Zehnder.
The one corresponds to path $b$ in Fig.\,\ref{fg:pmzprin}
is detected directly at Port~0.
The beam in path $a$ in Fig.\,\ref{fg:pmzprin}
is delivered to
HWP2 at an angle of $\pi/8$ and then to PBS3,
in order to visualize the interference of the beams
from the two paths.
The two outputs from PBS3 are detected at Ports 1 and~2.

The relative path length of the Mach--Zehnder
is adjusted to be a proper operating point
(which is the minimum at either of Port 1 or 2)
by a PZT actuator through a feedback system utilizing
the modulation-demodulation method.  
Once the relative path length is adjusted,
a sample-and-hold circuit keeps the mirror position fixed
during a measurement sequence (see below)
which lasts typically 20--30 seconds.

There are two photodetectors at each port,
a silicon photodiode and an APD
(avalanche photodiode, EG \& G, SPCM--AQ--141--FC)
guided through a multimode optical fiber.
The former is for alignment purpose (with increased light)
and the photon counting process is carried out with the latter,
by mechanically switching the beam between them.
The coupling efficiency of the fiber is measured to be
$0.75$--$0.8$, including the coupling lens and
the connectors before the APD\@.
The output from each APD is sent to a pulse counter
(EG \& G ORTEC, model 995) to count the number of
photon-induced pulses.

The counters are activated simultaneously by a common trigger,
typically of one-second duration and five-time repetition.
The numbers of counts in each duration are read
by a computer from all counters, so that we can
analyze statistical errors.
This procedure is repeated for each signal $\ketpsii$
with $i=0,\ldots,M{-}1$, composing a full sequence of
measuring the mutual information.
The ratio of counts in the three APDs provides
the channel matrix $P(y_j\vert x_i)$
from which the mutual information is derived.

As is discussed in Section~\ref{sc:sig$meas},
in the optimum detection scheme proposed,
the mutual information is increased by excluding
one of the possible signals.
Thus, realizing zero probabilities at the output ports
is essential in achieving a high mutual information.
In practice, however, there are several causes
that increase the probability at the output where
ideally zero is expected.
Among them, the most pronounced ones are
the pulses from an APD without any light (APD error),
the finite extinction ratio of a PBS (PBS error),
and the finite contrast of interference (interferometer error).

Without any light at all, the average dark counts of
the APDs were measured to be slightly less than 100\,count/sec.
Although the whole interferometer is enclosed in a box,
the environmental light increases the number of counts
to around 300\,count/sec,
even if no laser light is injected.
When the laser light is injected, the leak light due to
the imperfection of the interferometer is added,
and was measured to the average count of
around 1000\,count/sec for the output port
at which no count is expected ideally
(see Tables \ref{tb:TrineRatio}--\ref{tb:SeptenaryRatio3}).
The last increment is considered as the contributions
from the PBS errors and the interferometer error.
At the ports for which finite counts are expected, we had the
counts of order $10^5$\,count/sec at most,
which is within the linear range of APDs.

\begin{figure}
\begin{center}
\includegraphics[width=0.4\textwidth]{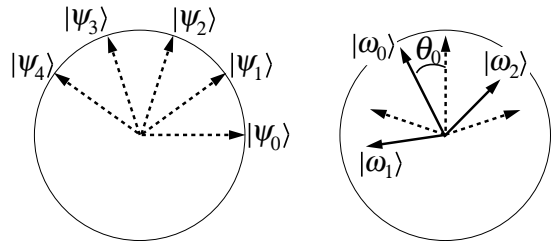}
\end{center}
\caption{\label{fg:5sig&w&w2}
The relation between the measurement state vectors
(left, quinary case $M=5$ in this example)
and the signal state vectors (right)
with an initial offset angle $\theta_0$.
}
\end{figure}

In general, a PBS has an angular-dependent separation
of two polarization components.
In our case, it turned out to be possible
to achieve the separation better than $1\,{:}\,1000$
for both polarization components,
by carefully aligning the angle of incidence
slightly ($\approx0.02$\,rad) different from
the standard value~$\pi/4$.
Then the expected contrast is ${}\approx0.998$,
which we thought sufficient for our experiment.
We adopted this angle in our polarization Mach--Zehnder
interferometer, resulting in a parallelogram arrangement
(see Fig.\,\ref{fg:layout}).

The actual contrast obtained with this interferometer
can be as high as
\begin{eqnarray}
{ { P_{\mathrm{max}}-P_{\mathrm{min}} }
  \over
  { P_{\mathrm{max}}+P_{\mathrm{min}} } }
\approx 0.98
\,,
\end{eqnarray}
though the typical values under normal experimental conditions
were slightly lower than this.
Thus, this is limited not by the PBS imperfection
but by, e.g., the spatial mode mismatch of the two beams.

In order to analyze the performance of our detecter circuit,
we measured not only the mutual information of the optimum
detection scheme but also its dependence on the relative
angle between the signal set $\{\,\ketpsii\}$
and the measurement state vectors $\{\,\ketomegaj\}$.
This is relevant to, for example, the possible
rotation of polarization in the transmitting fiber.
We measured the mutual information against
the signal set $\{\,\ketpsjithetao\}$ where
\begin{eqnarray}
\lefteqn{
\ketpsjithetao = {}
}\nonumber\\
&&
   \cos\!\left(\frac{i\pi}{M}+\theta_0\right) \ketH
 + \sin\!\left(\frac{i\pi}{M}+\theta_0\right) \ketV
\label{psidash_i}\\
&& (i=0, \ldots , M{-}1),
\nonumber
\end{eqnarray}
as a function of the initial offset angle $\theta_0$
(the optimum detection corresponds to $\theta_0=0$).
The relation between $\ketpsjithetao$ and
$\ketomegaj$ is depicted in Fig.\,\ref{fg:5sig&w&w2}
for the case of quinary signals.
In the experiment, $\theta_0$ was changed in steps of
$\pi/90$\ radian (two degrees).

%
%
\section{Results}
\label{sc:res}
%
\begin{figure}[t]
\begin{center}
\includegraphics[width=0.35\textwidth]{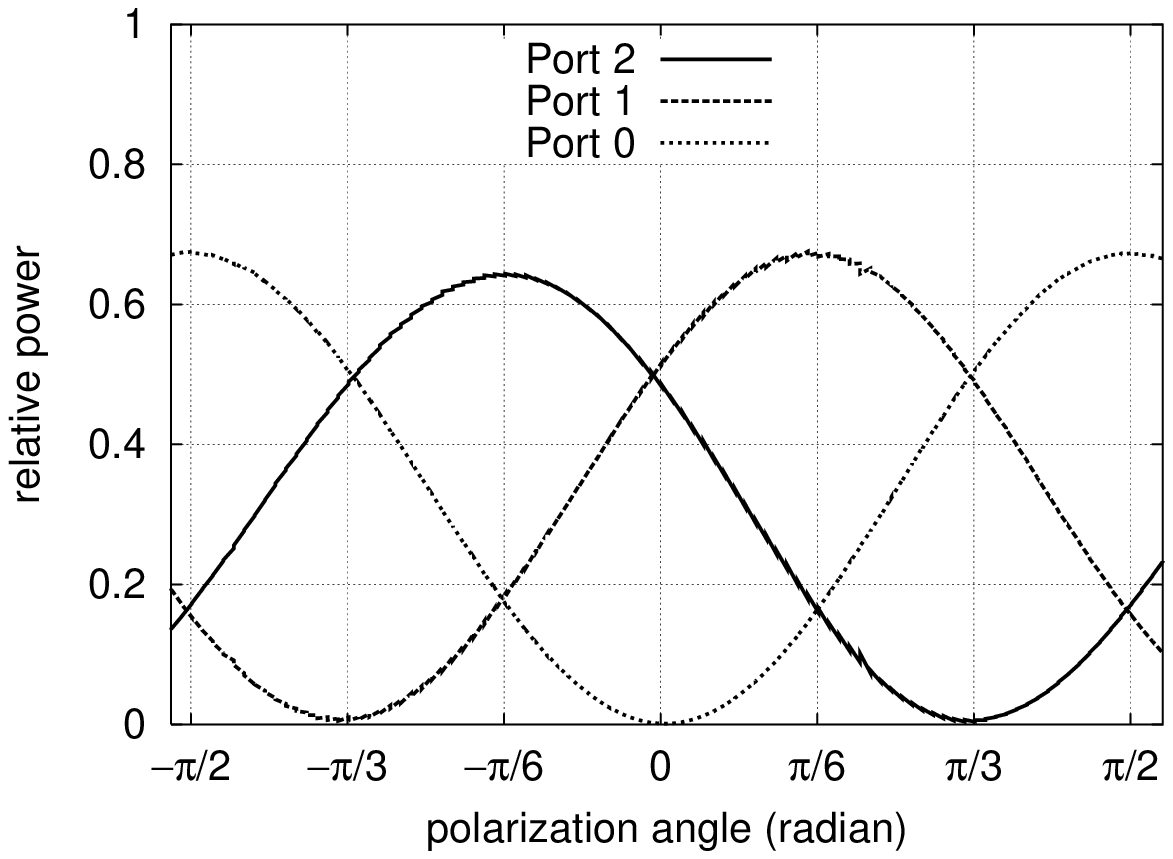}
\end{center}
\vspace{-1em}
\caption{\label{fg:TrineData}
The dependence of
the relative outputs at the three APDs
on the polarization angle of the injected beam
in the ternary experiment.
}
\begin{center}
\includegraphics[width=0.4\textwidth]{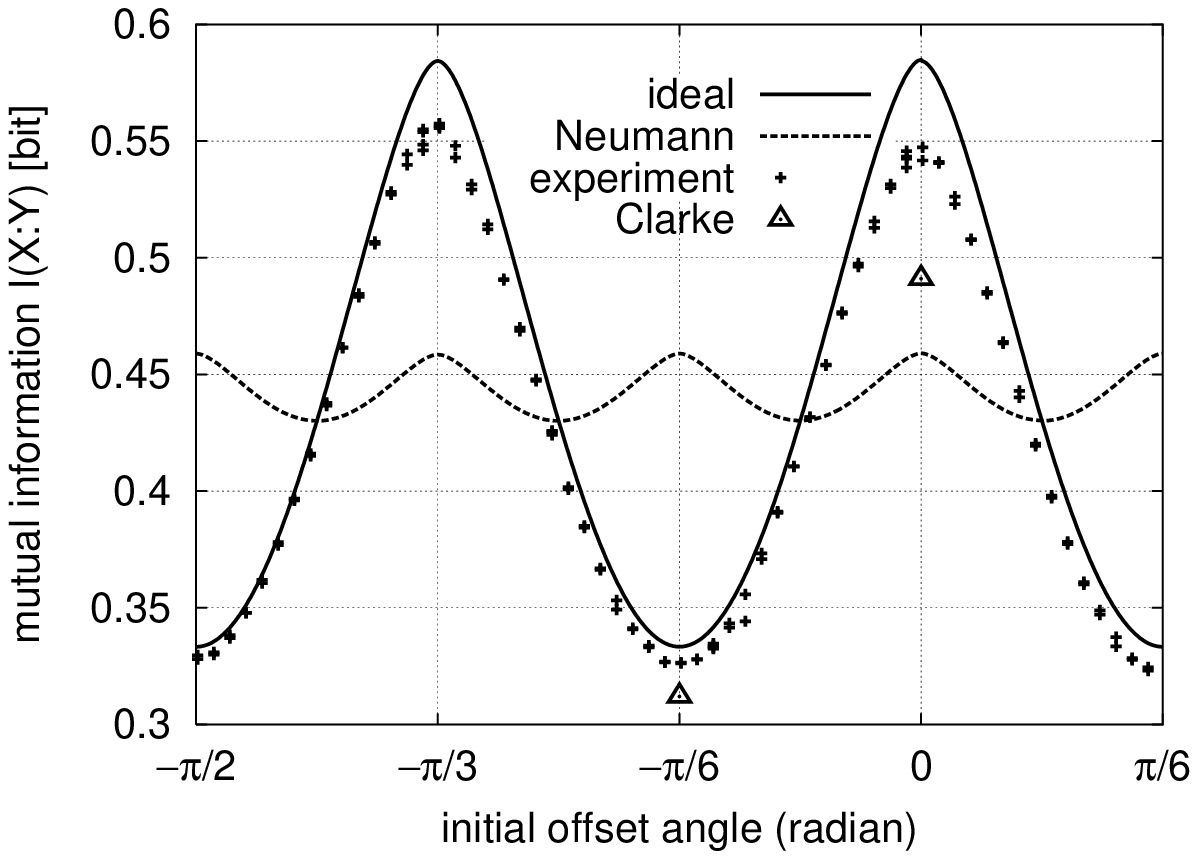}
\end{center}
\vspace{-1em}
\caption{\label{fg:TrineInf}
The dependence of the mutual information
on the initial offset angle $\theta_0$
in the ternary experiment (``experiment'', pluses).
The ideal case (``ideal'', solid curve) and
the ideal von Neumann case (``Neumann'', dashed curve)
are shown for comparison.
The values in an earlier experiment \cite{Clarke01b}
(``Clarke'', triangles at $\theta_0=0$ and $-\pi/6$)
are also shown.
}
\end{figure}

We carried out the optimum measurements described
in Section~\ref{sc:sig$meas} for the sources comprising
the ternary (trine), quinary and septenary states.
For the septenary signal states, both of the two optimum
detection schemes
(with $m=2$ and $m=3$ in Eq.\,(\ref{angle_gamma})\,)
were tested.

Fig.\,\ref{fg:TrineData} shows the relative output counts
at the three detectors
as the polarization of the input light is varied
in the ternary case.
This relative power corresponds to the probability for
the measurement outcome to occur for a single input photon.

\begin{figure}[t]
\begin{center}
\includegraphics[width=0.35\textwidth]{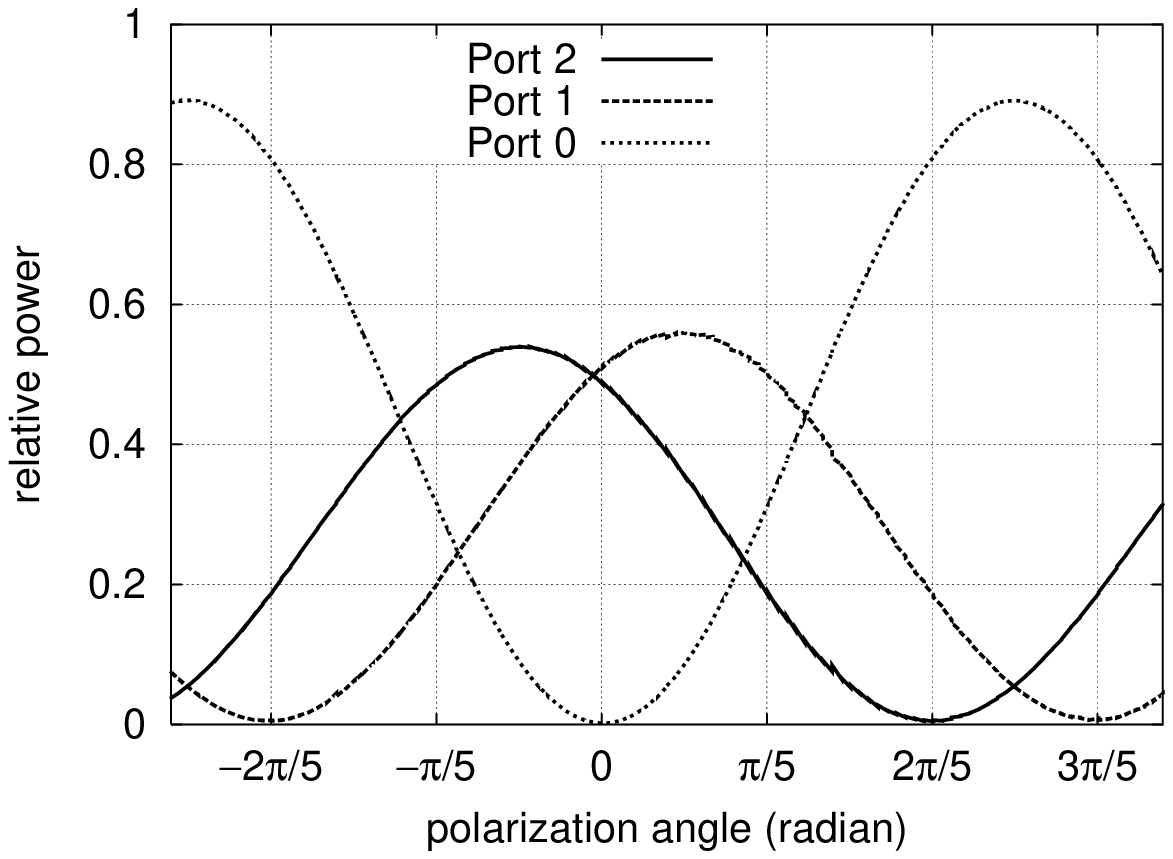}
\end{center}
\vspace{-1em}
\caption{\label{fg:QuinaryData}%
The dependence of
the relative outputs at the three APDs
on the polarization angle of the injected beam
in the quinary experiment.
}
\begin{center}
\includegraphics[width=0.4\textwidth]{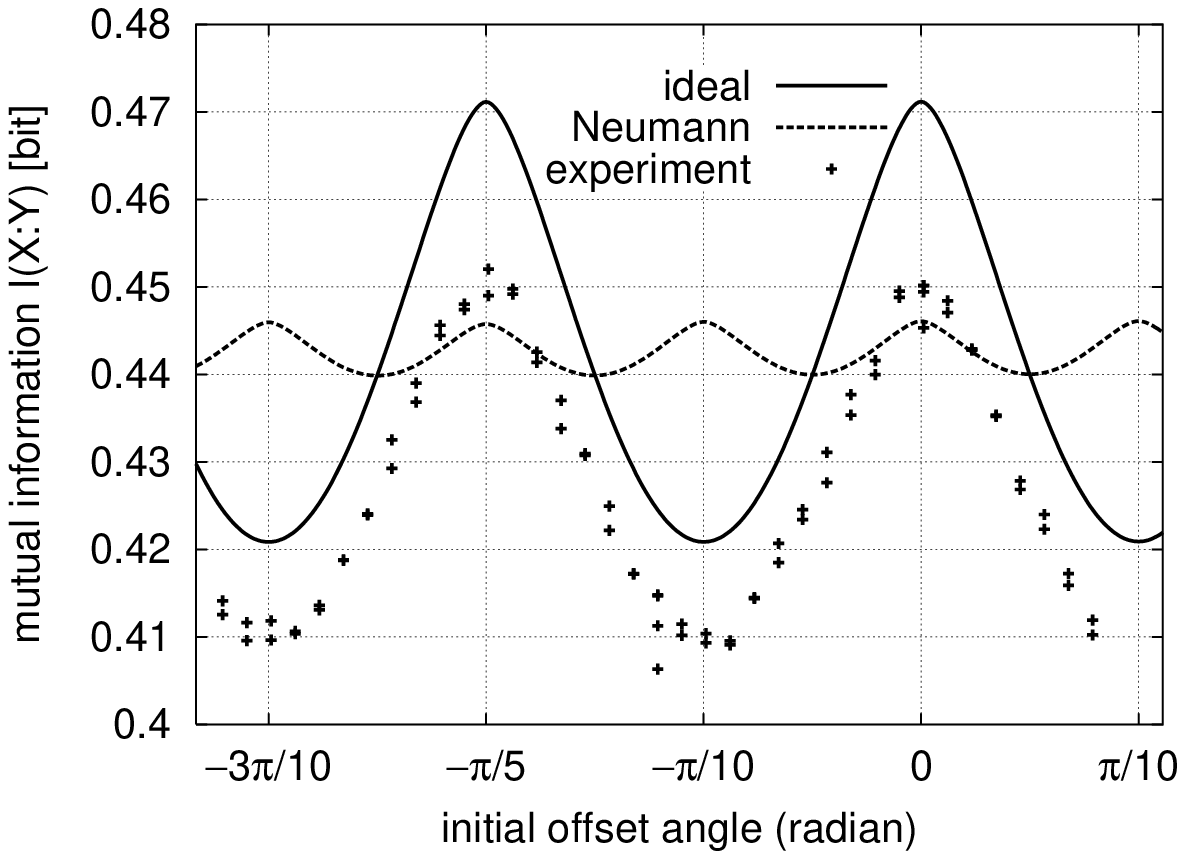}
\end{center}
\vspace{-1em}
\caption{\label{fg:QuinaryInf}
The dependence of the mutual information
on the initial offset angle $\theta_0$
in the quinary experiment.
The symbols are the same as in Fig.\,\ref{fg:TrineInf}.
}
\end{figure}

For the polarization angles $\{ -\pi/6, \pi/6, \pi/2 \}$
we are performing the state discrimination with the minimum error
probability, while for the angles $\{ -\pi/3, 0, \pi/3 \}$
we are realizing a measurement that allows unambiguous
elimination of one possibility among the three letter states.
These measurements were referred to as the trine and
anti-trine measurements in Ref.\,\cite{Clarke01b}.
These authors found an \emph{rms} deviation of $3.8\,\%$
from the theoretical value given in Table~\ref{tb:TrineRatio}.
Our results indicate a lower value of $1.1\,\%$.
The reason for our lower value is that we have been able to
achieve a smaller PBS error.

The data depicted in Fig.\,\ref{fg:TrineData} leads to
the mutual information presented in Fig.\,\ref{fg:TrineInf}.
At the optimum operating point, corresponding to the best
detection strategy, we clearly find that the mutual information
exceeds that attainable with the best von Neumann measurement.
Our value also exceeds that obtained earlier by
Clarke \textit{et al}.~\cite{Clarke01b}
represented as triangles in our figure.
The reason for this is again the smaller PBS error.
Our experimental value is slightly lower than
the theoretical maximum and this is due mainly to
a residual PBS error of approximately $0.1\,\%$
and also to the imperfect contrast of interference.
It was found \cite{pmzexp}
that despite the PBS error is not the
limiting factor of the interference contrast,
it has non-negligible effects on the mutual information.

Fig.\,\ref{fg:QuinaryData} shows the relative output counts
at our three detectors for the quinary case.
These provide the data with which to calculate
the mutual information depicted in Fig.\,\ref{fg:QuinaryInf}.
Our data show a marginal increase
in the mutual information beyond the value that may be
attained with the best von Neumann measurement.
The difference between our experimental result and
the theoretical value
is again principally attributable to the PBS error
and the imperfect contrast.

As mentioned earlier, the optimum detection scheme
increases the amount of the mutual information 
by excluding one of the possible signals.
With three detectors, only three signals can be
excluded at most, and the remaining signals do not
contribute the mutual information very much.
This fact reduces the maximum mutual information
in quinary case (and in septenary case as well)
from that in ternary case.
Although the absolute difference (of $\approx 0.02$)
between the experimental and ideal values
in the quinary case is similar to
that in the ternary case, the excess from the
von Neumann measurement became only marginal.

\begin{figure}
\begin{center}
\includegraphics[width=0.4\textwidth]{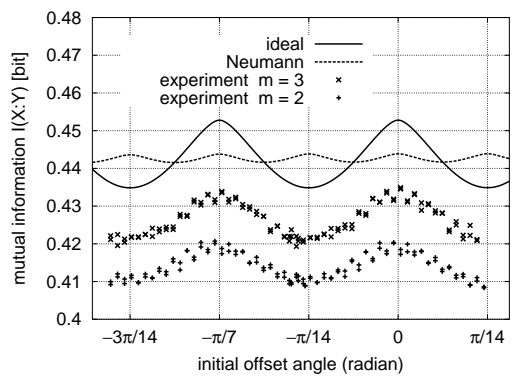}
\end{center}
\vspace{-1em}
\caption{\label{fg:SeptenaryData}
The dependence of the mutual information
on the initial offset angle $\theta_0$
in the septenary experiment with $m=2$ (crosses)
and $m=3$ (pluses).
Other symbols are the same as
in Fig.\,\ref{fg:TrineInf}.
}
\end{figure}

Fig.\,\ref{fg:SeptenaryData}
shows the mutual informations derived with
the two possible optimum detection schemes for the septenary case.
Even in an ideal case,
the increase in the attainable mutual information
over that found using the best von Neumann measurement is quite small.
In both cases our experimental values failed to reach even
the value attainable by means of the best von Neumann measurement.

The result with $m=3$ shows a higher mutual information
than that with $m=2$.
This difference is not by an experimental failure,
but due to the difference in the influences of
inperfect contrast between the two cases.
The reduced contrast increases the light leaking towards
the port where ideally no light is expected,
which in turn reduces the mutual information.
The absolute amount of leak light is proportional
to the amount of light interfering.
This qualitatively explains the difference of experimental
results with $m=2$ and $m=3$.
In the former case the interfering light is greater
than the latter, thus the influence on
the mutual information is larger.

%
%
\section{Discussion and concluding remarks}
Our ability to communicate classical information by means of a
quantum channel is limited by the existence of non-orthogonal
quantum states and the associated restrictions in discriminating
among them.  
These factors are fundamental to quanta as distinct
from classical information theory and make quantum key
distribution possible
\cite{PhoenixTownsend95,PhoenixBarnettChefles00}.

The optimum use of a quantum communication channel is closely
related to the maximization of mutual information,
as discussed in Appendix.    
The accessible information is obtained by
maximizing the mutual information through the selection
of the detection process.
There are only a very few examples of signal states
for which the accessible information is known
\cite{Levitin95_QCM94,Osaki2000_QCM98,Davies78,SasakiBarnettJozsa99}.
One such example is that of the real symmetric qubit states
\cite{SasakiBarnettJozsa99}.

In this paper we have described our polarization Mach--Zehnder
interferometer that was
designed to extract the accessible information from
signals formed from symmetric polarization states.
For the ternary (trine) states, our results proved
an amount of information close ($96\,\%$) to the theoretical limit.
Our value for the mutual information
exceeds that reported in an earlier experiment \cite{Clarke01b}.
The difference between our measured value for the mutual information
and the theoretical limit is due principally to the leakage of the
`wrong' polarization through our polarizing beam splitters
and also to the imperfect contrast.
The effect of this leakage is more pronounced
when we consider the quinary and septenary signal states.
Our experiments suggest that
optimum quantum communication based on the ternary (trine)
polarization states,
for example the quantum key distribution by
the Phoenix--Barnett--Chefles protocol
\cite{PhoenixBarnettChefles00},
should be feasible.
Schemes based on the quinary
and septenary states will present a greater challenge.

In the light of fundamental interests,
the quinary and septenary states meet with the simplest cases
where the maximum amount of information can be extracted by
a detection in which the number of possible outputs is
less than that of input states.
Davies' theorem predicted that a device with three possible outputs
suffices for any real polarization system of a single photon.
In our experiment, Davies' theorem has been tested
within the PBS error.
For the complete confirmation, further study might be
necessary, e.g.\ comparing the minimum-output optimum detection
with the one corresponding the group covariant optimal solution
which consists of the same number of outputs as inputs.

%
%
\begin{acknowledgments}
We are grateful to Dr.~Izutsu, Professor Hirota,
Dr.~Riis, and Dr.~Clarke
for discussion and encouragement.
This work was supported, in part, by the British Council,
the Royal Society of Edinburgh, and by the Scottish
Executive Education and Lifelong Learning Department.
\end{acknowledgments}
%
%
\appendix*
\section{Mutual information}
\label{sc:MutInf}
In this Appendix we give the definition of
the mutual information and explain its functional meaning.
Primary concerns of information theory are
\textit{how to represent messages as effectively as possible}
and
\textit{how to transmit messages as precisely as possible}.
The mutual information is related with the second problem.

A sender has a source of messages $S$ and selects one of
a known set $\{a, b, \ldots, z\}$
with given prior probabilities
$\{ P(a), P(b), \ldots, P(z) \}$.
This source may be characterized by the random variable
$S=\{ a, b, \ldots, z; P(a), P(b), \ldots, P(z) \}$.
The sender represents each of these messages by
a sequence of a given set of letters
$\{x_i\}$ such as $\{0,1\}$.
These are the symbols running through the transmission channel.
Each message is then represented by a codeword
formed from a sequence of letters.
This is \textit{source coding}.
Information theory tells us that
the effectiveness of source coding can be measured
by the minimum of the average length required for
a codeword and that it is given by the Shannon entropy
\begin{equation}
H(S) = -\sum_{\mskip-100mu A=a,b,\ldots\mskip-100mu}
   P(A) \log_2 P(A)
\,.
\label{H(S)}
\end{equation}
This is a measure of uncertainty in the random variable $S$.
It takes its maximum value when all elements
appear with equal probability, that is, when we know
nothing better than a random guess for each element.
This measure of uncertainty is regarded as the amount of information
required to represent $S$.

A channel is usually subject to various types of noise disturbances.
Information theory provides means and limits for reliable information
transmission with such noisy channels.
The key idea is to introduce some redundancy in the codeword
representation prior to transmission so as to allow the correction
of errors at the receiving side.
This entails adding some redundant
letters to the codewords and hence increases their length.
This is \textit{channel coding}.
The mutual information quantifies how much redundancy is required
for error-free transmission.

The output from the source encoder is a sequence of the letters
forming the codewords representing the messages.
For such sequences one can find the frequencies of appearance
$P(x_i)$ for each letter $x_i$.
Thus we can define a random variable $X=\{ x_i ; P(x_i) \}$ for
the outputs from the source encoder.
This is the set of inputs to the channel.
A mathematical model for the channel is specified by the set of
possible outputs $\{ y_j \}$ and the conditional probability
$P(y_j\vert x_i)$ for each input.
Given $X$, $\{ y_j \}$, and $[P(y_j\vert x_i)]$,
we can determine the existence or nonexistence of encoders and decoders
that achieve a given level of transmission performance.

The mutual information is defined between the input and output
random variables $X$ and $Y=\{ y_j ; P(y_j) \}$.
Here
\begin{equation}
P(y_j)\equiv\sum_{x_i} P(y_j\vert x_i)P(x_i)
\label{P(y)}
\end{equation}
is the probability of having $y_j$.
The uncertainty of the input random variable $X$
is measured by the Shannon entropy
\begin{equation}
H(X)=-\sum_i P(x_i) \log P(x_i)
\label{H(X)}
\end{equation}
defined in a similar way to Eq.\,(\ref{H(S)}).

If the receiver detects the output signal $y_j$,
then he is now more certain about $X$.
The new probability distribution conditioned by $y_j$ is given as
\begin{equation}
P(x_i\vert y_j)=\frac{P(y_j\vert x_i)P(x_i)}{P(y_j)}
\,.
\label{BayesRule}
\end{equation}
One can then define the average conditional entropy by
\begin{equation}
H(X\vert Y)=-\!\sum_{y_j}\! P(y_j)
             \!\sum_{x_i} P(x_i\vert y_j) \log P(x_i\vert y_j)
.
\label{H(X|Y)}
\end{equation}
This quantifies the remaining uncertainty of $X$ after having
the knowledge on the conditioning variable $Y$.
The information extracted by the receiver
is naturally defined by the reduction of the uncertainty,
\begin{eqnarray}
\lefteqn{
\IXY = H(X)-H(X|Y)
}\nonumber\\
&=&
\sum_{x_i, y_j}
    P(x_i)P(y_j\vert x_i)
    \log \biggl[ { P(y_j\vert x_i) \over
                  {\displaystyle\mathop{\textstyle\sum}_{x_i}
                         P(x_i)P(y_j\vert x_i)}}
    \biggr]
\,.
\label{IXY}
\end{eqnarray}
This $\IXY$ is the mutual information between $X$ and~$Y$.

Now let us consider a block coding of length~$n$.
The output from the source encoder is a letter sequence,
which is devided into blocks (\emph{message blocks})
of length $k$ (${<}n$).
Each block is supplemented by an additional block
(\emph{correction block}) of $n{-}k$ letters
to compose a transmission codeword $\{\mathbf{x}_p\}$\,:
\begin{eqnarray}
\mathbf{x}^p &=&
\overbrace{{x^p}_{1}{x^p}_{2}\cdots {x^p}_{k}}%
   ^{\displaystyle\text{message block}}\,
\overbrace{{x^p}_{k+1}{x^p}_{k+2}\cdots {x^p}_{n}}%
   ^{\displaystyle\text{correction block}}
\\
&&\text{(for  $p=1,2,\ldots,L^k$)}
\,,
\nonumber
\end{eqnarray}
where each ${x^p}_{l}$ $(l=1,\cdots,n)$ is an element
of possible letters $\{ x_i; i=0,1,\cdots,L{-}1 \}$.
Note that although there are $L^n$ possible sequences
of length~$n$ in total,
only part of them, i.e.\ $L^k$ sequences,
are used as codewords.
This redundancy,
together with appropriate choice of correction blocks,
allows us to recover the possible errors in transmission.

The input codeword $\mathbf{x}^p$ will be disturbed
in the channel so as to come out as a different sequence
$\mathbf{y}^q = {y^q}_{1}{y^q}_{2}\cdots {y^q}_{n}$.
The channel decoder processes this output codeword
to assign an appropriate sequence
which should be the correct input codeword.
The average error in this decoding
should be as small as possible, while the redundancy
$n-k$ should also be as small as possible.
In other words, keeping the ratio $R=k/n$,
so-called the \emph{transmission rate}, as large as possible,
we wish to attain a small error in decoding.

Let us suppose that encoding is made under the constraint that
the frequency of $x_i$'s occurring in the set of codewords
$\{\mathbf{x}^p\}$ is $P(x_i)$.
Information theory says that
by an appropriate design of the coding scheme
it is possible to transmit the messages
with an error probability as small as desired
if $R<\IXY$ is satisfied.
For the fixed channel model $[P(y_j\vert x_i)]$, one may further
adjust prior probabilities $\{ P(x_i) \}$ to maximize the mutual
information.
The maximum value
\begin{equation}
C_{\mathrm{c}}=\max_{\{P(x_i)\}}\IXY
\end{equation}
is called the \emph{channel capacity}.
Then the channel coding theorem tells us
\cite{Shannon48,Gallager_book,Cover&Thomas_book}
that if $R<C$ holds
there exists a coding scheme which transmits messages
with an error probability as small as desired.
Thus the mutual information is related to
the ultimate use of the channel.

The basic frameworks described above
also apply to a quantum limited channel.
However a new ingredient comes into play, which is a quantum
effect in the detection process.
Let us consider the simplest case where the letter set $\{x_i\}$
is conveyed by a set of pure states $\{\,\ketpsii\}$, possibly
a nonorthogonal set, through a noiseless channel.
Then the channel model is specified by a POM $\{\widehat\Pi_j\}$
and the channel matrix
$P(y_j\vert x_i) = \brapsii \widehat\Pi_j \ketpsii$.
The POM describes the measurement process and gives it a
quantum prescription for generating the output letters $\{y_j\}$.

In the conventional (classical) context,
the channel matrix $[P(y_j\vert x_i)]$ is given and fixed.
In quantum domain, however, one may ask what is
the best possible POM for the given set of letter states
$\{\,\ketpsii\}$.
This is actually a nontrivial problem as discussed
in Introduction.
The problem can be decomposed into several steps.
First we can consider the maximization of the mutual information
with respect to a POM $\{\widehat\Pi_j\}$ for the fixed
$\{\,\ketpsii\}$
and prior probabilities $\{ P(x_i) \}$.
The maximum value
\begin{equation}
I_{\mathrm{Acc}} \bigl(\{ \ketpsii ; P(x_i) \}\bigr)
= \max_{\{\widehat\Pi_j\}}
  I \bigl(\,\{ \ketpsii ; P(x_i)\} \,{:}\,Y \bigr)
\end{equation}
is called the \textit{accessible information} of
$\{ \ketpsii ; P(x_i) \}$.
We can then consider the maximization of the accessible
information over prior probabilities $\{ P(x_i) \}$, and may
define the quantity $C_1$ as
\begin{equation}
C_1 = \max_{\{P(x_i)\}}
    I_{\mathrm{Acc}} \bigl(\,\{ \ketpsii ; P(x_i)\,\} \bigr).
\end{equation}
This would be a natural extension from the conventional idea.
However, this $C_1$ is not in general the maximum bound for
the transmission rate for error-free communication, and hence
it is not the channel capacity.
In fact, there is the peculiar quantum interference effect
in quantum detection of codeword states, which was not
taken into account in the conventional theory.
The true capacity for a pure state channel is given
by Hausladen \textit{et al.}~\cite{Hausladen96_coding}.
The general theory for a mixed state channel is given by
Holevo~\cite{Holevo96_coding1} and by
Schumacher and  Westmoreland~\cite{Schumacher97_coding}.

To realize reliable transmission ensured by quantum theory
of the channel capacity, one may need quantum computation
for the decoding process~\cite{Sasaki98a,Sasaki98_realization}.
This is, however, far beyond present
technologies.
If only a quantum detection on each letter state is available,
then $I_{\mathrm{Acc}}$ and $C_1$ practically specify the limit of
communication ability.
Let us suppose again that encoding is made such that
$x_i$ (i.e.\ $\ketpsii$) occurs in the set of codewords
$\{\mathbf{x}^p\}$
 (i.e.\ $\{\,\ketpsiio\otimes\cdots\otimes\ketpsiin \}$)
with the probability $P(x_i)$.
We further suppose that $\{\widehat\Pi_j\}$ is the POM attaining the
accessible information for $X=\{ \ketpsii ; P(x_i) \}$
and the receiver applies this detection \textit{separately} on each
letter states to get output sequences
$\{y_{j_1}y_{j_2}\cdots y_{j_n}\}$.
If $R<I_{\mathrm{Acc}}$ holds, then a reliable transmission of
the letters with an arbitrarily small error is possible
by an appropriate \textit{classical} coding.
The optimum POM for the accessible information is thus an
important concern for devising a good code for a quantum
limited channel.

%
%
%

%
%
\end{document}